\documentclass[lettersize,journal]{IEEEtran}
\usepackage{amsmath,amsfonts}
\usepackage{algorithmic}
\usepackage{array}
\usepackage[caption=false,font=normalsize,labelfont=sf,textfont=sf]{subfig}
\usepackage{textcomp}
\usepackage{stfloats}
\usepackage{url}
\usepackage{verbatim}
\usepackage{graphicx}
\usepackage{tabularx,booktabs}
\usepackage{cite}
\usepackage{bm}
\usepackage{array}
\usepackage{graphicx}
\usepackage[justification=centering]{caption}
\usepackage{algorithm}
\usepackage{amssymb}
\usepackage{algorithmic}
\usepackage[T1]{fontenc}
\usepackage{tabularx}
\usepackage{subfig}
\def\BibTeX{{\rm B\kern-.05em{\sc i\kern-.025em b}\kern-.08em
    T\kern-.1667em\lower.7ex\hbox{E}\kern-.125emX}}
\usepackage{balance}
\begin{document}

\title{Low-Communication Resilient Distributed Estimation Algorithm Based on Memory Mechanism}
\author{Wei~Li, Limei~Hu, Feng~Chen,~\IEEEmembership{Member,~IEEE,} Ye~Yao
\thanks{Wei Li, Limei Hu and Feng Chen are with the College of Artificial Intelligence, Southwest University, Chongqing 400715, China, and also with the Key Laboratory of Luminescence Analysis and Molecular Sensing, Southwest University, Ministry of Education, Chongqing 400715, China (e-mail: weili9805@163.com; hlm0903@swu.edu.cn; fengchen.uestc@gmail.com).
Ye Yao is with the School of Electronic Information and Communications, Huazhong University of Science and Technology, Wuhan 430074, China(e-mail: yeyao@hust.edu.cn) (Corresponding author: Feng Chen.)
}}

\markboth{}%
{How to Use the IEEEtran \LaTeX \ Templates}

\maketitle

\begin{abstract}
In multi-task adversarial networks, the accurate estimation of unknown parameters in a distributed algorithm is hindered by attacked nodes or links. To tackle this challenge, this brief proposes a low-communication resilient distributed estimation algorithm. First, a node selection strategy based on reputation is introduced that allows nodes to communicate with more reliable subset of neighbors. Subsequently, to discern trustworthy intermediate estimates, the Weighted Support Vector Data Description (W-SVDD) model is employed to train the memory data. This trained model contributes to reinforce the resilience of the distributed estimation process against the impact of attacked nodes or links.  Additionally, an event-triggered mechanism is introduced to minimize ineffective updates to the W-SVDD model, and a suitable threshold is derived based on assumptions. The convergence of the algorithm is analyzed. Finally, simulation results demonstrate that the proposed algorithm achieves superior performance with less communication cost compared to other algorithms.

\end{abstract}


\section{Introduction}
\IEEEPARstart{W}{ith} the recent advancements in wireless communication technology, distributed estimation has garnered significant attention\cite{ref1}, which plays a crucial role in various domains, including Wireless Sensor Networks (WSNs), Internet of Things (IoT) networks and smart grids\cite{ref2, Shao, ref3,ref4,10752873, 10848372}.

In recent years, many distributed estimation algorithms have been proposed\cite{ref5,ref6,ref7,ref8,ref9,ref10} and they perform excellently in normal network conditions. However, the real-world wireless sensor networks are not always secure, as they can easily be affected by environmental factors or targeted attacks\cite{ref11,ref12}. In order to remove or reduce the impact of these insecure nodes on network performance, a series of distributed secure estimation algorithms have emerged.

In the adversarial network environment with Byzantine attack, a distributed algorithm adopting local minswitching decision (LMSD) is proposed in \cite{ref13} to counteract compromised nodes or links. a secure diffusion LMS algorithm based on the hybrid-subsystem (DLMS-H) is proposed in \cite{ref14}, which can effectively handle Byzantine and manipulation attacks.
In the adversarial network environment with False Data Injection (FDI) attack, a secure diffusion LMS (S-DLMS) algorithm composed of two subsystems as presented in \cite{ref15} can detect the nodes under attacks. Additionally, a two-step resilient distributed estimation method is proposed in \cite{ref16}, which is capable of breaking the traditional assumption that the number of attacked nodes is less than half of the node neighborhood. Similarly, the adaptive width generalized correntropy diffusion algorithm (AWGC-DA) proposed in \cite{ref17} also breaks the aforementioned traditional assumption and exhibits excellent performance. It is worth noting that many distributed security algorithms have also been proposed in multitask networks \cite{ref18,ref19,ref20}. The main focus is on FDI and link attacks in the network in this brief.

On the other hand, algorithms based on diffusion strategy require all nodes to send and receive intermediate estimates from their neighbors. This additional inter-node communication consumes more limited resources including energy and bandwidth, thereby affecting the lifespan of sensors. Therefore, reducing network communication costs is also a concern in the field of distributed estimation. The reduced-communication diffusion least mean-square (RC-DLMS) algorithm proposed by \cite{ref21} adopts a strategy of randomly selecting communication nodes, which reduces communication costs in a secure network environment. However, in the adversarial network environments, this method can easily incur communication costs when communicating with attacked nodes. An intelligent selection method that aims to reduce communication costs by refusing to communicate with nodes prone to pulse noise interference is introduced in \cite{ref22}. Similar to the latter, this brief proposes a node selection strategy based on reputation value to reduce communication costs.

In summary, this brief proposes a low-communication resilient distributed estimation algorithm in the adversarial multitask network with node and link attacks. The main contributions are as follows:

\noindent1) A resilient distributed estimation algorithm based on memory mechanism is proposed, which utilizes memory data for training W-SVDD model to detect anomalous data. Additionally, an event-trigger mechanism is introduced to prevent performance degradation resulting from model overfitting.

\noindent2) A node selection method based on reputation value is proposed, which allows partially more secure neighbors to send information to the node and achieves the purpose of reducing communication cost.

\noindent3) Convergence analysis and simulation results demonstrate that the proposed algorithm can effectively handle FDI and link attacks in the network with smaller communication cost.

\section{Preliminaries and problem formulation}
\subsection{Weighted Support Vector Data Description(W-SVDD)}
The Weighted Support Vector Data Description (W-SVDD) introduced in\cite{ref23} is a widely used anomaly detection algorithm. The core idea of this algorithm is to construct a hypersphere that tightly envelops the sample data to detect anomalous data.
Given a dataset $\boldsymbol{X}=\{\boldsymbol{x_i}\}_{i=1}^{v}\in{{\Re}^{{L}\times{v}}}$, the optimization problem for W-SVDD can be expressed as:
\begin{equation}
\begin{aligned}
     \text{min} & \quad  {{r}^2}+{P}{\sum\limits_{i=1}\limits^{v}}{b_i}{\xi_i} &\\
     \text{s.t.}& \quad\|{\Phi(\boldsymbol x_i)}-{\boldsymbol o}\|^2\leq{r^2}+{\xi_i},\text{ }{\xi_i}\geq 0,\text{ }{{i}={1,2,...,v}}\text{,} &
    \label{eq.W-SVDD0}
\end{aligned}
\end{equation}
where ${\boldsymbol o}$ and ${r}$ represent the center and radius of the hypersphere, the slack variable ${\xi_i}$ allows a small amount of data outside the hypersphere, ${P}$ is the penalty parameter, ${b_i}$ denotes the weight assigned to the $i$-th sample, the feature map ${\Phi}$ helps transform data from the original space to the specified feature space.

To avoid explicitly computing inner products in high-dimensional space, the original problem ({\ref{eq.W-SVDD0}}) can be transformed into an equivalent dual problem by using the Lagrange multiplier method for solution:
\begin{equation}
\begin{aligned}
     \text{max }& \quad{\sum\limits_{i=1}\limits^{v}} {\alpha_i}{K}({\boldsymbol x_i},{\boldsymbol x_i})-{\sum\limits_{i=1}\limits^{v}} {\sum\limits_{j=1}\limits^{v}} {\alpha_i}{\alpha_j}{K}({\boldsymbol x_i},{\boldsymbol x_j})&\\ 
     \text{s.t. }& \quad {\sum\limits_{i=1}\limits^{v}}{\alpha_i}=1,&\\
     & \quad 0\leq{\alpha_i}\leq{b_i}{P},\text{ }{{i}={1,2,...v}}{\text{,}}&\label{eq.dualProblem}
\end{aligned}
\end{equation}
where ${\alpha_i}$ represents the Lagrange multiplier for the $i$-th sample. and the Gaussian kernel function ${K({\boldsymbol x_i},{\boldsymbol x_j})}={\langle{{\Phi}({\boldsymbol x_i}),{\Phi}({\boldsymbol x_j})}\rangle}={exp}^{(-{\gamma}{\|{\boldsymbol x_i}-{\boldsymbol x_j}\|}^2)}$ is introduced to compute inner products in high-dimensional space.

The Lagrange multiplier vector ${\boldsymbol \alpha}={\{\alpha_1,\alpha_2,...,\alpha_i,\alpha_v\}}$ can be obtained by solving problem ({\ref{eq.dualProblem}}). The radius ${r}$ is defined as the distance from the support vector on the boundary to the center of the sphere, as shown below:
\begin{equation}
    {r}=\sqrt{{K}({\boldsymbol x_{s}},{\boldsymbol x_{s}})-2{\sum\limits_{i=1}\limits^{v}}{\alpha_i}{K}({\boldsymbol x_{s}},{\boldsymbol x_i})+{\sum\limits_{i=1}\limits^{v}} {\sum\limits_{j=1}\limits^{v}} {\alpha_i}{\alpha_j}{K}({\boldsymbol x_i},{\boldsymbol x_j})}\text{,}
\end{equation}
where ${\boldsymbol x_s}\in\{{\boldsymbol x_{i}}| 0<\alpha_{i}< b_{i}P,i=1,2...,v\}$ represents the support vector. The decision function ${\mathcal{F}(\boldsymbol x_p)}$ for the data ${\boldsymbol x_p}$ can be obtained by comparing the distance from ${\boldsymbol x_p}$ to the center of the sphere and the radius $r$:
\begin{equation}
\begin{aligned}
    {\mathcal{F}(\boldsymbol x_p)}={K}({\boldsymbol x_{p}},{\boldsymbol x_{p}})-2{\sum\limits_{i=1}\limits^{v}}{\alpha_i}{K}({\boldsymbol x_{p}},{\boldsymbol x_i})&\\+{\sum\limits_{i=1}\limits^{v}} {\sum\limits_{j=1}\limits^{v}} {\alpha_i}{\alpha_j}{K}({\boldsymbol x_i},{\boldsymbol x_j})-{r^2}.
\end{aligned}
\end{equation}
The data ${\boldsymbol x_p}$ is considered normal if ${\mathcal{F}(\boldsymbol x_p)}\leq0$ and classified as an outlier otherwise.

\subsection{Multi-task DLMS}
Consider a multi-task  network with $N$ nodes, where the $N$ nodes are divided into $M$ clusters.
Each node $n$ estimates an unknown parameter ${\boldsymbol {w}_{n}^o}$ by utilizing a set of gathered data $\{d_{n,t},{\boldsymbol u_{n,t}}\}$ at time $t$, where $d_{n,t}$ and ${\boldsymbol u_{n,t}}$ respectively signify the real-valued scalar measurement and $L$-dimensional regression vector. The following linear relationship is satisfied:
\begin{equation}
{{d}_{n,t}}=\boldsymbol u_{n,t}^{T}\boldsymbol w_{n}^{o}+{{z}_{n,t}},
\label{data model}
\end{equation}
where ${\boldsymbol w_{n}^{o}}$ is an unknown $L$-dimensional column vector and ${{z}_{n,t}}$ is the zero-mean measurement noise. In the multi-task network, node $n$ and $l$ within the same cluster share the same estimation target while node $n$ and $k$ from different clusters have different but somewhat similar estimation targets, as follows:
\begin{equation}
\left\{ \begin{aligned}
& \boldsymbol w_{n}^{o} ={\ }\boldsymbol w_{l}^{o},{\forall n,l} \in {{\mathcal{C}_n}} \\ 
& \boldsymbol w_{{n}}^{o} \sim{\ }\boldsymbol w_{k}^{o},\text{  if   }{{\mathcal{C}_n}\neq{\mathcal{C}_k}, }
\end{aligned} \right.
\end{equation}
where ${{\mathcal{C}_n}}$ denotes the cluster to which node $n$ belongs and $
\sim$ represents similarity

For the DLMS algorithm, the Adaptive-Then-Combine (ATC) strategy is adopted due to its superior performance compared to the Combine-Then-Adaptive (CTA) strategy. In multitask networks, the clustered multitask DLMS algorithm proposed in \cite{ref3} utilizes the similarity between different tasks to further enhance network performance, as illustrated below:

\begin{equation}
\left\{ \begin{aligned}
& {{\boldsymbol \psi }_{n,t+1}}={{\boldsymbol w}_{n,t}} +\mu (\sum\limits_{l\in {{\mathcal{N}}_{n}}\cap  {\mathcal{C}_n}}{{{c}_{ln}}({{d}_{l,t}}-\boldsymbol u_{l,t}^{T}\boldsymbol w_{n,t}^{{}}})\boldsymbol u_{l,t}\\ 
& \text{}+\mu\eta {\sum\limits_{k\in {{\mathcal{N}}_{n}}\backslash  {\mathcal{C}_n}}}{\rho}_{kn}({{\boldsymbol w}_{k,t}}-{{\boldsymbol w}_{n,t}}))\text{    }\text{               (Adapt)},  \\ 
& {{\boldsymbol w}_{n,t+1}}=\sum\limits_{l\in {{\mathcal{N}}_{n}}\cap  {\mathcal{C}_n}}{{{a}_{ln}}{{\boldsymbol \psi }_{ln,t+1}}}\text{                (Combine)}, 
\label{eq.MDLMS}
\end{aligned} \right.
\end{equation}
where ${\mathcal{N}_n}$ denotes the neighboring nodes including the node $n$ itself. ${{\boldsymbol \psi }_{n,t+1}}$ and ${{\boldsymbol \psi }_{ln,t+1}}$ are the intermediate estimate computed by node $n$ and transmitted by the same-cluster neighbor $l$ at time $t+1$, respectively. $\mu$ and $\eta$ stands for the step size and regularization parameter. ${{a}_{ln}}$ and ${{c}_{ln}}$ are the fusion coefficients for the adaptive and combination steps from $l$ to $n$. ${\rho}_{kn}$ is the contribution from the non-same-cluster neighbor $k$ to $n$. For convenience in subsequent discussion, let ${\mathcal{N}_{n}^{+}}$ and ${\mathcal{N}_{n}^{-}}$ denote the same-cluster neighbors and the non-same-cluster neighbors of node $n$ respectively, i.e. ${\mathcal{N}_{n}^{+}}={{\mathcal{N}}_{n}}\cap  {\mathcal{C}_n}$ and ${\mathcal{N}_{n}^{-}}={{\mathcal{N}}_{n}}\backslash  {\mathcal{C}_n}$

\subsection{Attack Models}
In this brief, False Data Injection (FDI) attack and link attack models are considered.

\noindent1)	\emph {FDI Attack}: The attacker misleads the estimation of the target ${\boldsymbol{w}_{n}^{o}}$ by tampering with the measurement ${d_{n,t}}$ of node $n$ through data injection:
\begin{equation}
    d_{n,t}=\left\{\begin{aligned}
    &\boldsymbol u_{n,t}^{T}{\boldsymbol w_{n}^{o}}+{{z}_{n,t}} \text{,     Node }n\text{ is secure}\text{;} \\
    &\boldsymbol u_{n,t}^{T}({\boldsymbol w_{n}^{o}+{\boldsymbol w_{n,t}^{att} }})+{{z}_{n,t}}\text{,     Node }n\text{ is attacked}\text{,}
    \end{aligned}\right.
\end{equation}
where ${\boldsymbol w_{n,t}^{att} }$ is an $L$-dimensional random vector. 

\noindent2)	\emph {Link Attack}: The attacker manipulates the communication link between node $n$ and its neighbor $l$ to make $n$ receive incorrect intermediate estimate.
\begin{equation}
    {{\boldsymbol \psi }_{ln,t}}=
    \left\{\begin{aligned}
     &{\boldsymbol \psi_{l,t}} \text{,   Link }ln\text{ is secure}\text{;} \\ 
    &{\boldsymbol \psi_{l,t}}+{\boldsymbol \psi_{l,t}^{att}} \text{,   Link }ln\text{ is attacked}\text{,} 
    \end{aligned}\right.
\end{equation}
where ${n} \in {{{\mathcal{N}}_{l}}}\backslash\{l\}$ and ${\boldsymbol \psi_{l,t}^{att} }$ is an $L$-dimensional vector.

To design a multi-task distributed resilient estimation algorithm, the following assumption is made:

$A1$: For each node $n$, the number of neighbor nodes that are under attack is less than half of all its neighbor nodes. This assumption is common in related works.

\section{The proposed resilient M-DLMS algorithm based on memory mechanism}
In this section, a resilient distributed estimation algorithm based on a memory mechanism is proposed. The proposed algorithm comprises five parts: Adaption, Node selection, Communication, Detection, and Combination.

The node $n$ undergoes the following steps to update ${\boldsymbol w_{{n,t+1}}}$ at time $t+1$:

\noindent1)	Adaption: Node $n$ exchanges estimated values with non-same-cluster neighbors $k, {k \in {\mathcal{N}}_{n}^{-}} $ and update its own intermediate estimate $\boldsymbol \psi_{n,t+1}$ by using information through the following expression:
\begin{equation}
\begin{aligned}
& {{\boldsymbol \psi }_{n,t+1}}={{\boldsymbol w}_{n,t}}+\mu ({{d}_{n,t}}-\boldsymbol u_{n,t}^{T}\boldsymbol w_{n,t}^{{}})\boldsymbol u_{n,t}\\ 
& \text{}+\mu\eta {\sum\limits_{k\in {\mathcal{N}}_{n}^{-}}{\rho}_{nk}({{\boldsymbol w}_{k,t}}-{{\boldsymbol w}_{n,t}}))} \\ 
\end{aligned}.
\label{Adaption}
\end{equation}


\noindent2) Communication: Node $n$ receives intermediate estimate from the neighbor $l_{1}, {l_{1} \in { {\mathcal{N}}_{n}^{+}}}$ and sends its own intermediate estimate to the neighbor $l_{2}, {n \in { {\mathcal{N}}_{n}^{+}}}$.

\noindent3) Detection: Node $n$ utilizes the data in memory and the received intermediate estimate to solve the optimization problem for W-SVDD and trains a detection model. The model categorizes the same-cluster neighbor node into either the secure node set ${S_{n,t+1}^{+}}$ or the insecure node set ${S_{n,t+1}^{-}}$ and updates memory data simultaneously.

\noindent4) Combination: The nodes in set ${S_{n,t+1}^{+}}$ should have equal importance in updating $\boldsymbol{w}_{n,t+1}$. Therefore, the fusion coefficients are determined by using average fusion strategy in this brief:
\begin{equation}
    {{a}_{ln}}=\left\{\begin{aligned}
    &\frac{1}{|{S}_{n,t+1}^{+}|} \text{,   if   }l \in {S_{n,t+1}^{+}}\text{;}\\
    &0\text{,               if  }l \notin {S_{n,t+1}^{+}}\text{,}\\
    \end{aligned}\right.
\end{equation}
node $n$ updates ${{\boldsymbol w}_{n,t+1}}$ using the following expression:
\begin{equation}
    {{\boldsymbol w}_{n,t+1}}=\sum\limits_{l\in {{\mathcal{N}}_{n}}\cap  {\mathcal{C}_n}}{{{a}_{ln}}{{\boldsymbol \psi }_{ln,t+1}}}.
\end{equation}
Next, detailed explanations will be provided.

\subsection{Detection}
The node $n$ obtains the intermediate estimate from its same-cluster neighbors after the communication phase, denoted by ${\boldsymbol \Psi_{{n,t+1}}}=\{{\boldsymbol \psi_{{l_1n,t+1}}},{\boldsymbol \psi_{{l_2n,t+1}}},...,{\boldsymbol \psi_{{l_{|{ {\mathcal{N}}_{n}^{+}}|}n,t+1}}}\}$, where $l_1,l_2,...,l_{{{\mathcal{N}}_{n}^{+}}} \in {{\mathcal{N}}_{n}^{+}}$. Moreover, certain historical secure data is stored in the memory of the node $n$, as shown below:
\begin{equation}
    {\boldsymbol \Omega_{n,t+1}}=\{\boldsymbol \Psi_{S_{n,t-le}^{+}},\boldsymbol \Psi_{S_{n,t-(le-1)}^{+}},...,\boldsymbol \Psi_{S_{n,t}^{+}}\},
\label{history data}
\end{equation}
where $le$ is the length of the sliding window and ${\boldsymbol \Psi_{S_{n,t}^{+}}}$ is defined as follow:
\begin{equation}
    {\boldsymbol \Psi_{S_{n,t}^{+}}}=\{{\boldsymbol \psi_{{s_1n,t}}},{\boldsymbol \psi_{{s_2n,t}}},...,{\boldsymbol \psi_{{s_{|S_{n,t}^{+}|}n,t}}}\}, s_1,s_2,...,s_{|S_{n,t}^{+}|} \in S_{n,t}^{+},
\label{secure intermediate estimate}
\end{equation}
where $S_{n,t}^{+}$ represents the safe neighbors detected by node $n$ at time $t$.
Then the trustworthy neighbors of node $n$ at time $t+1$ can be detected by executing the following steps:

\noindent1) Dataset construction: Node $n$ constructs a training set by using the secure data in memory ${\boldsymbol \Omega_{n,t}}$ and the received intermediate estimate $\boldsymbol \Psi_{{n,t+1}}$: 
\begin{equation}
    {\boldsymbol \varPsi_{n,t+1}}=\{{\boldsymbol \Omega_{n,t}},\boldsymbol \Psi_{{n,t+1}}\}.
\label{train data}
\end{equation}
and subsequently centralizing the training set:
\begin{equation}
    {\boldsymbol \varPsi_{n,t+1}^{mean}}={\boldsymbol \varPsi_{n,t+1}}-\text{mean}{({\boldsymbol \Omega_{n,t})},}
\label{mean train data}
\end{equation}
where mean$(\cdot)$ denotes the operation of taking the average value. It is worth noting that the centering operation in this brief using only the average of the historical safe data ${\boldsymbol \Omega_{n,t}}$ is to avoid the adverse effects of potentially anomalous data in $\boldsymbol \Psi_{{n,t+1}}$.

\noindent2) Training: The Lagrange multiplier vector ${\boldsymbol \alpha_{n,t+1}}$ and decision function ${\mathcal{F}_{n,t+1}(\cdot)}$ are obtained by solving the W-SVDD optimization problem Eq. (\ref{eq.W-SVDD0}) for the dataset ${{\boldsymbol \varPsi_{n,t+1}^{mean}}}$.

\noindent3) Determination: Node $l \in {{\mathcal{N}}_{n}^{+}}$ is categorized into the set of secure nodes ${S_{n,t+1}^{+}}$ or the set of attacked nodes ${S_{n,t+1}^{-}}$ based on the detection result, as follows:
\begin{equation}
l\in\left\{ \begin{aligned}
&   {S_{n,t+1}^{+}}  \text{,  if   } {\mathcal{F}_{n,t+1}(\boldsymbol\psi_{ln,t+1}}-\text{mean}{({\boldsymbol \Omega_{n,t})})\leq 0}\text{;}\\ 
&  {S_{n,t+1}^{-}}  \text{,  if   } {\mathcal{F}_{n,t+1}(\boldsymbol\psi_{ln,t+1}}-\text{mean}{({\boldsymbol \Omega_{n,t})})> 0}\text{.}
\end{aligned} \right.
\label{deteciont result}
\end{equation}
\noindent4) Memory data update: Node $n$ updates the historical safe ${\boldsymbol \Omega_{n,t+1}}$ in memory by using and Eq. (\ref{history data}) respectively.

\subsection{Updating Model Based on Event-triggering Mechanism}
Up to now, each node $n$ retrains the W-SVDD model to detect anomalous data at each time. However, such practices are computationally wasteful and may even result in a decline in algorithm performance due to overfitting. To address these issues, an event-triggered mechanism is introduced to ensure necessary updates to the model.

Denote $y_{n,d}$ as the $d$-th event-triggered moment for node $n$ to update the model. The sequence of triggering instants are defined as the set ${\{y_{n,d}|y_{n,d}\in\mathbb{N}_{+}\}}$ where $0<y_{n,0}<y_{n,1}<...<y_{n,d}$. Before the next model update, i.e. $y_{n,d} < t+1 < y_{n,d+1}$, the average value for centralization, mean${({\boldsymbol \Omega_{n,y_{n,d}})}}$ and the decision function ${\mathcal{F}_{n,y_{n,d}}}(\cdot)$ are stored in node $n$.




Diverging from the Detection delineated in subsection A, the node $n$ first detects the anomalous data at time $t+1$ by using the existing model upon receiving intermediate estimate from its neighbors, as follows:
\begin{equation}
l\in\left\{ \begin{aligned}
&   {S_{n,t+1}^{+}}  \text{,  if   } {\mathcal{F}_{n,y_{n,d}}(\boldsymbol\psi_{ln,t+1}}-\text{mean}{({\boldsymbol \Omega_{n,y_{n,d}})})\leq 0}\text{;}\\ 
&  {S_{n,t+1}^{-}}  \text{,  if   } {\mathcal{F}_{n,y_{n,d}}(\boldsymbol\psi_{ln,t+1}}-\text{mean}{({\boldsymbol \Omega_{n,y_{n,d}})})> 0}\text{.}
\end{aligned} \right.
\label{ ETM deteciont result}
\end{equation}
It indicates that the model can still achieve the desired performance without the need for retraining when the number of anomalous nodes $|{S_{n,t+1}^{-}}|$ falls below the threshold $\theta_{n}$, i.e. $|{S_{n,t+1}^{-}}|\leq\theta_{n}$.
Then the node $n$ skips steps \noindent1) - \noindent3) outlined in subsection A and directly execute the step \noindent4) to complete the detection process.

Otherwise, if the number of anomalous nodes $|{S_{n,t+1}^{-}}|$ exceeds the threshold $\theta_{n}$, i.e. $|{S_{n,t+1}^{-}}|\geq\theta_{n}$, it indicates that the current model is no longer suitable for the current environment and should be updated immediately. Then the node $n$ will fully execute steps \noindent1) - \noindent4) as described in subsection A to complete the detection process. The next trigger update  can be performed using the following equation:
\begin{equation}
y_{n,d+1}=y_{n,d}+\{{t+1|{\text{ }|{S_{n,t+1}^{-}}|}\geq{\theta_n},\}    }
\end{equation}
The next issue to consider is how to choose an appropriate threshold $\theta_{n}$. According to assumption $A1$, it can be easily inferred that the threshold should be set as follows:
\begin{equation}
\theta_{n}=\frac{|{{\mathcal{N}}_{n}^{+}}|  }{2}
\end{equation}

\section{The proposed algorithm with low communication}

In general, the methods based on diffusion strategies have higher communication loads. Taking node $n$ as an example, it exchanges data with all same-cluster neighbors in each communication phase, which generates a total $L|{{\mathcal{N}}_{n}^{+}}|$ communication load. When the bandwidth resources of node $n$ are not enough to support such a large amount of communication, problems such as data loss may occur and thereby impacting algorithm performance. To address this challenge, this section proposes a node selection method based on reputation values. This method allows node $n$ to exchange information only with selected neighbors, effectively reducing the data communication load.

Before introducing the proposed node selection method, it is necessary to explain some additional steps. Firstly, node $n$ assigns an evaluation value $\lambda_{nl,t}$ to same-cluster neighbor $l$ at time $t$ based on the result after completing the detection process, as shown below:
\begin{equation}
    \lambda_{nl,t}=\left\{\begin{aligned}
     &1  \text{,     if    } l\in{S_{n,t}^{+}},\\ 
     &0  \text{,     if    } l\in{B_{n,t}^{-}},\\
     &-1  \text{,    if   } l\in{S_{n,t}^{-}},\\
    \end{aligned}\right.
\end{equation}
where ${S_{n,t}^{+}},{B_{n,t}^{-}}$ and ${S_{n,t}^{-}}$ respectively represent secure neighbors, neighbors not chosen for communication and unsafe neighbors for node $n$ at time $t$, they are pairwise disjoint and satisfy the following relationships:
\begin{equation}
    \begin{aligned}
    &{B_{n,t}^{+}}={S_{n,t}^{+}}\cup{S_{n,t}^{-}},\\
    &{\mathcal{N}}_{n}^{+}={B_{n,t}^{+}}\cup{B_{n,t}^{-}}.
    \end{aligned}
\end{equation}

After the adaption based on Eq. (\ref{Adaption}) at time $t+1$, the node $n$ calculates the reputation value of each same-cluster neighbor $l$ as follows:
\begin{equation}
\begin{aligned}
    {{\tau}_{{nl},t+1}}=\sum\limits_{i=t-le}^{t}{{{\lambda}_{nl,i}},
    }
\end{aligned}
\end{equation}
When the reputation value ${{\tau}_{{nl},t+1}}$ is larger, it means that node $l$ has behaved more securely in the recent time and is more trustworthy to node $n$. Conversely, when ${{\tau}_{{nl},t+1}}$ is less than 0, it means that node $l$ is more insecure and node $n$ should refuse to choose it as a communication neighbor.
Then node $n$ ranks the reputation values of its neighbors in descending order:
\begin{equation}
\begin{aligned}
    {\boldsymbol{\tau}_{{n},t+1}}={\{{\tau}_{n{l_1},t+1},{\tau}_{n{l_2},t+1},...{\tau}_{n{l_{|{\mathcal{N}_{n}^{+}}|-1}},t+1}\}},
\end{aligned}
\end{equation}
where ${\tau}_{n{l_1},t+1}\geq{\tau}_{n{l_2},t+1}\geq...\geq{\tau}_{n{l_{|{\mathcal{N}_{n}^{+}}|-1}},t+1}$ and $l_1,l_2,...,l_{|{\mathcal{N}_{n}^{+}}|-1}$ are the corresponding neighbor nodes. The node $n$ selects the top $i$ nodes with the highest reputation values to form communication set $B_{n,t+1}^{+}$, and ${p_{n}}=\frac{i}{|{B_{n,t+1}}|}$ is the communication node ratio. The remaining unselected nodes form the set $B_{n,t+1}^{-}$. Then the node $n$ communicates with the selected neighbor $l$, $l\in {B_{n,t+1}^{+}}$.

\section{Performance analysis}
In this section, the performance analysis of the proposed algorithm is presented and the following assumption is adopted:

\emph{Assumption 1}: The input vector ${\boldsymbol u_{n,t}}$ is independent across both time and space  with zero mean and covariance matrix ${R_{u,n}}=E\{{\boldsymbol u_{n,t}}{\boldsymbol u_{n,t}^T}\}>0$.

The weight error vector of node $n$ at the moment $t$ is defined as:
\begin{equation}
    \begin{aligned}
        \Tilde{\boldsymbol{w}}_{n,t}=\boldsymbol{w}_{n,t}-\boldsymbol{w}_{n}^{o},
    \end{aligned}
    \label{error estimation}
\end{equation}
In order to facilitate the analysis of the estimation performance of the whole network, the block vectors corresponding to the weight error vector $\Tilde{\boldsymbol{w}}_{n,t}$, the weight estimate vector ${\boldsymbol{w}}_{n,t}$, the ideal weight vector ${\boldsymbol{w}}^{o}_{n}$, and the intermediate estimate vector ${\boldsymbol{\psi}}_{n,t+1}$ at time $t$ are defined as:
\begin{equation}
\begin{aligned}
    \Tilde{\boldsymbol{w}}_{t+1}&=col\{\Tilde{\boldsymbol{w}}_{n,t+1}\}_{n=1}^{N},
\end{aligned}
\label{eq.block weight error vector}
\end{equation}
\begin{equation}
\begin{aligned}
    {\boldsymbol{w}}_{t+1}&=col\{{\boldsymbol{w}}_{n,t+1}\}_{n=1}^{N},
\end{aligned}
\label{eq.block weight estimate vector}
\end{equation}
\begin{equation}
\begin{aligned}
    {\boldsymbol{w}}^{o}&=col\{{\boldsymbol{w}}_{n}^{o}\}_{n=1}^{N},\\
\end{aligned}
\label{eq.block ideal weight vector}
\end{equation}
\begin{equation}
\begin{aligned}
    {\boldsymbol{\psi}}_{t+1}&=col\{{\boldsymbol{\psi}}_{n,t+1}\}_{n=1}^{N}.
\end{aligned}
\label{eq.block intermediate estimate}
\end{equation}

Define the local error vector $\overline{\boldsymbol{w}}_{n,t}\triangleq\boldsymbol{w}_{n,t}-\boldsymbol{w}^{o}$, the global vector $\overline{\boldsymbol{w}}_{t}\triangleq col\{\overline{\boldsymbol {w}}_{n,t}\}^{N}_{k=1}$, $\boldsymbol{\psi}_{t}\triangleq col\{\boldsymbol {\psi}_{n,t}\}^{N}_{k=1}$. By subtracting $\boldsymbol {w}^{o}$ from both sides of the first equation in equation (\ref{eq.MDLMS}), the following can be obtained:
\begin{equation}
\begin{aligned}
{{ \boldsymbol \psi }_{t+1}}-{{ \boldsymbol w}^{o}}=&{\overline{\boldsymbol{w}}_{t}}-\mu {{ \boldsymbol H}_{u,t}}{\overline{\boldsymbol{w}}_{t}}-\mu \eta \boldsymbol Q({\overline{\boldsymbol{w}}_{t}}+{{ \boldsymbol w}^{o}}\text{)}\\
&+\mu {{ \boldsymbol p}_{zu,t}} +\mu { \boldsymbol{\delta_t}}  ,
\end{aligned}
\label{eq.errorBlock}
\end{equation}
where 
\begin{equation}
\begin{aligned}
{{ \boldsymbol \Omega}_{u,t}}&=diag\{\sum\limits_{l\in {{\mathcal{N}}_{n}}\cap  {\mathcal{C}_n}}{{{c}_{ln}}{{ \boldsymbol u}_{l,t}} \boldsymbol u_{l,t}^{T}}\}_{n=1}^{N},\\ 
{ \boldsymbol{\delta_t}}&=col\{\sum\limits_{l\in {{\mathcal{N}}_{n}}\cap  {\mathcal{C}_n}}{{{c}_{ln}}{{ \boldsymbol u}_{l,t}} \boldsymbol u_{l,t}^{T}{{ \boldsymbol w_{n,t}^{att} }}}\}_{n=1}^{N}, \\
{{ \boldsymbol p}_{zu,t}}&=col\{\sum\limits_{l\in {{\mathcal{N}}_{n}}\cap  {\mathcal{C}_n}}{{{c}_{ln}} \boldsymbol u_{l,t}^{T}{{z}_{l.t}}}\}_{n=1}^{N},\\
\boldsymbol Q&={{ \boldsymbol I}_{LN}}- \boldsymbol G\otimes {{I}_{L}},
\end{aligned}
\end{equation}
where $\otimes$ represents the Kronecker product and $\boldsymbol G$ denotes an N$\times$N matrix with $(n, l)$-th entry $\rho_{nl}$, and $\boldsymbol G_{nn}=1$ if $n$ has only same-cluster neighbors.

Let ${{ \boldsymbol A}_{I}}= \boldsymbol A\otimes {{ \boldsymbol I}_{L}}$. Similarly, subtract $\boldsymbol {w}^{o}$ from both sides of Eq. (\ref{eq.MDLMS}) for the second term, substitute Eq. (\ref{eq.errorBlock}), and then taking the expectation of both sides yields:
\begin{equation}
\begin{aligned}
E\{\overline{\boldsymbol{w}}_{t+1}\}&= \boldsymbol A_{I}^{T}[{{ \boldsymbol I}_{LN}}-\mu ({{ \boldsymbol \Omega}_{R}}+\mu  \boldsymbol Q)]E\{\overline{\boldsymbol{w}}_{t}\}\\
&-\mu \eta  \boldsymbol A_{I}^{T} \boldsymbol Q{{w}^{o}}
+\mu  \boldsymbol A_{I}^{T}E\{{ \boldsymbol{\delta_t}}\}+\boldsymbol A_{I}^{T}E\{{{\boldsymbol \delta^{*}_t}}\},
\label{eq.expec}
\end{aligned}
\end{equation}
where ${{\boldsymbol \delta^{*}_t}}=col\{\boldsymbol {\psi_{n,t}^{att}}\}^{N}_{n=1}$, ${{ \boldsymbol \Omega}_{R}}=E\{{{ \boldsymbol \Omega}_{u,i}}\}=diag\{{{ \boldsymbol R}_{1}},{{ \boldsymbol R}_{2}},...{{ \boldsymbol R}_{N}}\}$ and ${{ \boldsymbol R}_{n}}=\sum\limits_{l\in {{\mathcal{N}}_{n}}\cap  {\mathcal{C}_n}}{{{c}_{ln}}E\{{{ \boldsymbol u}_{n,i}} \boldsymbol u_{n,i}^{T}\}}$. When all attacks are detected, $E\{{ \boldsymbol{\delta_t}}\}=0$ and $E\{{{\boldsymbol \delta^{*}_t}}\}=0$, the above expression becomes as follows:
\begin{equation}
\begin{aligned}
E\{\overline{\boldsymbol{w}}_{t+1}\}&= \boldsymbol A_{I}^{T}[{{ \boldsymbol I}_{LN}}-\mu ({{ \boldsymbol \Omega}_{R}}+\mu  \boldsymbol Q)]E\{\overline{\boldsymbol{w}}_{t}\}-\mu \eta  \boldsymbol A_{I}^{T} \boldsymbol Q{{w}^{o}}.
\label{eq.expec2}
\end{aligned}
\end{equation}
The proposed algorithm asymptotically converges to the mean when the step size satisfies the following conditions\cite{ref3}:
\begin{equation}
0<\mu <\frac{2}{{{\max }_{n}}\{{{\lambda }_{\max }}({{ \boldsymbol R}_{n}})\}+2\eta }.
\label{mu}
\end{equation}
The asymptotic mean deviation is
\begin{equation}
\begin{aligned}
\underset{t\to \infty }{\mathop{\lim }}\,E\{\overline{\boldsymbol{w}}_{t}\}&=\mu \eta {{\{ \boldsymbol A_{I}^{T}[{{ \boldsymbol I}_{LN}}-\mu ({{ \boldsymbol \Omega}_{R}}+\eta  \boldsymbol Q)]-{{I}_{LN}}\}}^{-1}}\boldsymbol A_{I}^{T} \boldsymbol Q{{ \boldsymbol w}^{o}}.
\label{54}
\end{aligned}
\end{equation}

\section{Simulation results}
\begin{figure}[t]
	\centering
	\subfloat[]{
		\includegraphics[width=0.34\textwidth]{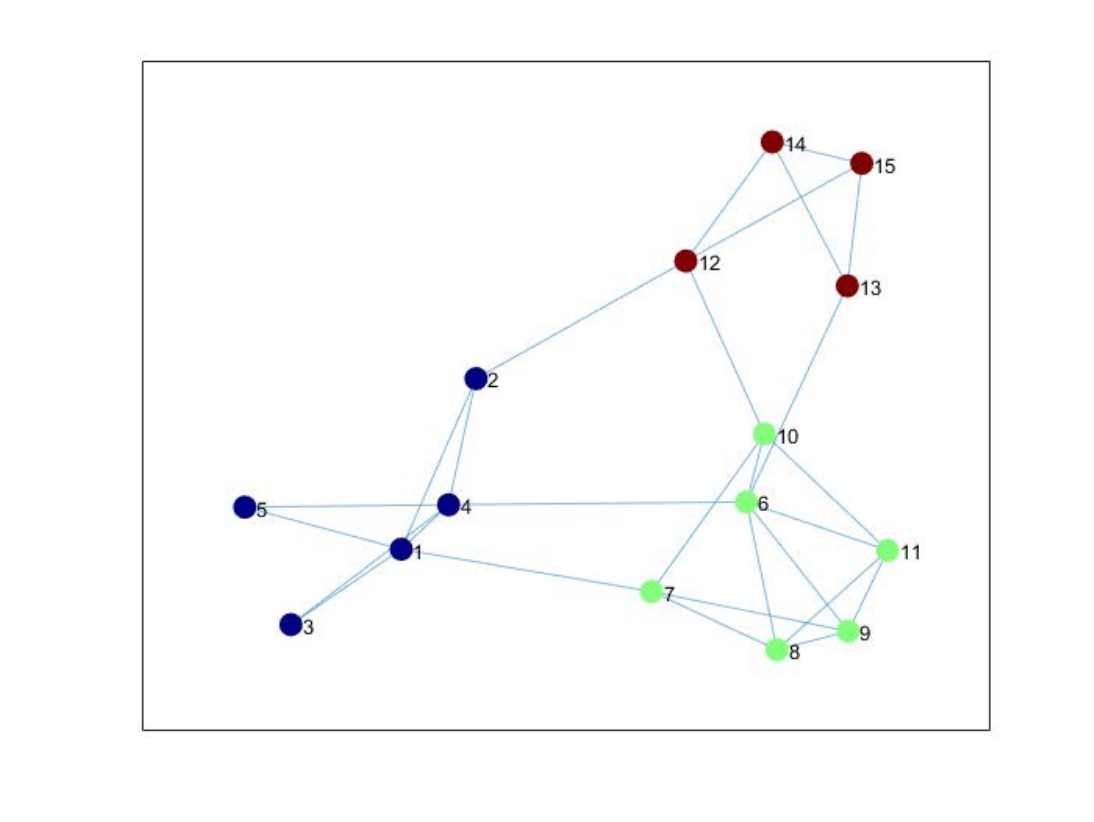}
	}\hspace{0.1in}%
	\subfloat[]{
		\includegraphics[width=0.34\textwidth]{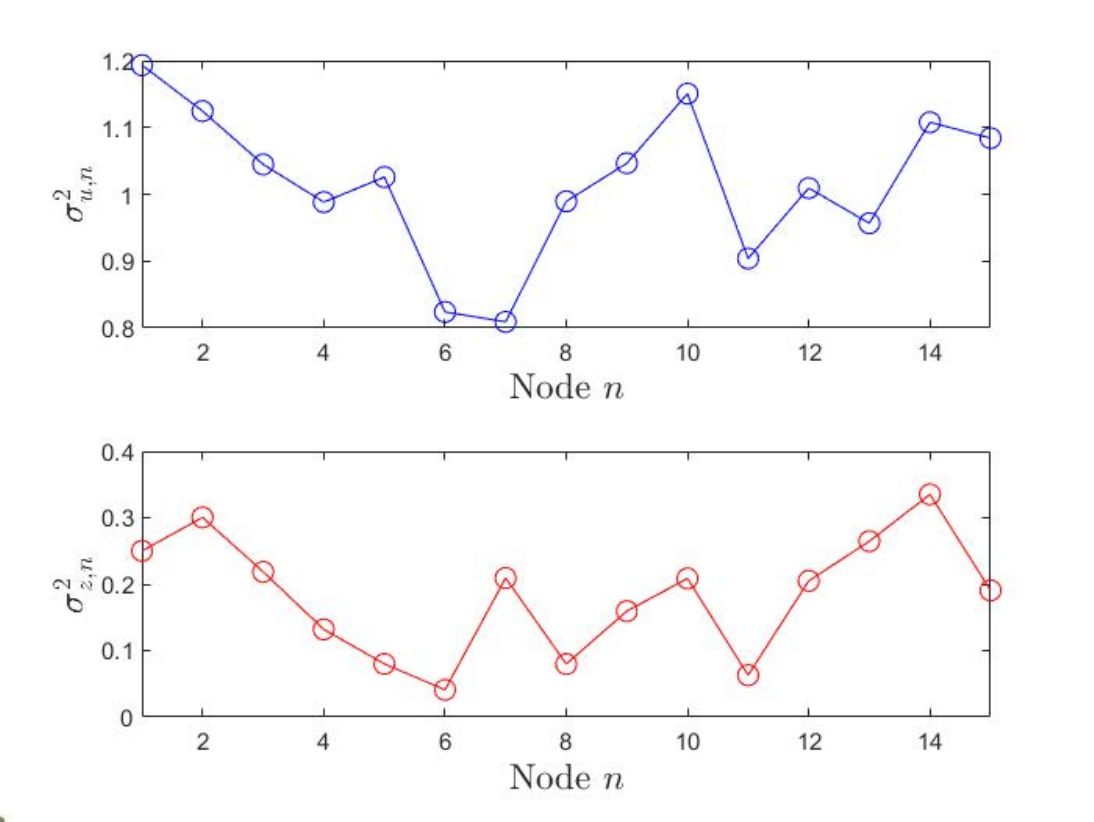}
	}
	\caption{Multi-task network topology and signal and noise variances.
		(a) Multi-task network topology.
		(b) Input signal variances (top) and noise variances (bottom).}
	\label{network_model}
\end{figure}

As shown in Fig. \ref{network_model} (a), a multi-task network consisting of 15 nodes is considered in this brief, and the nodes are divided into three clusters: $C_{1}=\{1,2,3,4,5\}$, $C_{2}=\{6,7,8,9,10,11\}$, $C_{3}=\{12,13,14,15\}$. The regression input vector $\boldsymbol {u}_{n,t}$ is a three-dimensional zero-mean Gaussian random vector with a covariance matrix $\boldsymbol {R}_{u,n}={\sigma}^2_{u,n}{\boldsymbol {I}_{L}}$, and the noise ${z}_{n,t}$ is a zero-mean Gaussian random variable with a covariance matrix $\boldsymbol {R}={\sigma}^{2}_{z,n}{\boldsymbol I}_{L}$. The signal variance ${\sigma}^2_{u,n}$ and the noise variance ${\sigma}^{2}_{z,n}$ are shown in Fig. \ref{network model}(b).
The simulation results are obtained by conducting 200 independent Monte Carlo experiments. The $MSD_t={10\log_{10}}\frac{1}{N}{\sum}^{N}_{n=1}||{\boldsymbol w^o-\boldsymbol w_{n,t}}||^{2}_{2}$ is used as a performance metric. 

The parameter settings are as follows: The dimension of an unknown vector $L=3$, step size $\mu= 0.03$, regularization parameter $\eta=0.02$, sliding window length $le=2$, penalty
parameter $P=0.4$, kernel radius $\gamma=50$. For convenience, the weight of secure data is uniformly set to $0.9$, and the weight of received data is uniformly set to $0.4$. The attack values $\boldsymbol w_{n,t}^{att}$ and ${\boldsymbol \psi}_{n,t}^{att}$ generated by Gaussian distributions $N(0,3)$ and $N(0,0.5)$ respectively. The proposed algorithm compares with NC-LMS, M-DLMS \cite{ref3}, S-DLMS \cite{ref18}, AWGC-DA \cite{ref17} with $\beta=1$ algorithms. 
It should be noted that the GC-DA algorithm cannot defend against link attacks. Therefore, this algorithm is not included in the comparison during experiments involving link attacks.

\renewcommand{\arraystretch}{1}
\begin{table}
  \centering
  \caption{AVERAGE NUMBERS OF EVENTS FOR EACH NODE}
  \begin{tabular}{|c|c|}
    \hline
    Communication node ratio & Average number ofevents \\
    \hline
    1 & 72.5 \\
    \hline
    1/2 & 89.6 \\
    \hline
    1/3 & 244.6 \\
    \hline
  \end{tabular}
  \label{table1}
\end{table}

\begin{figure}[t]
	\centering
	\subfloat[FDI attack]{%
		\includegraphics[width=0.34\textwidth]{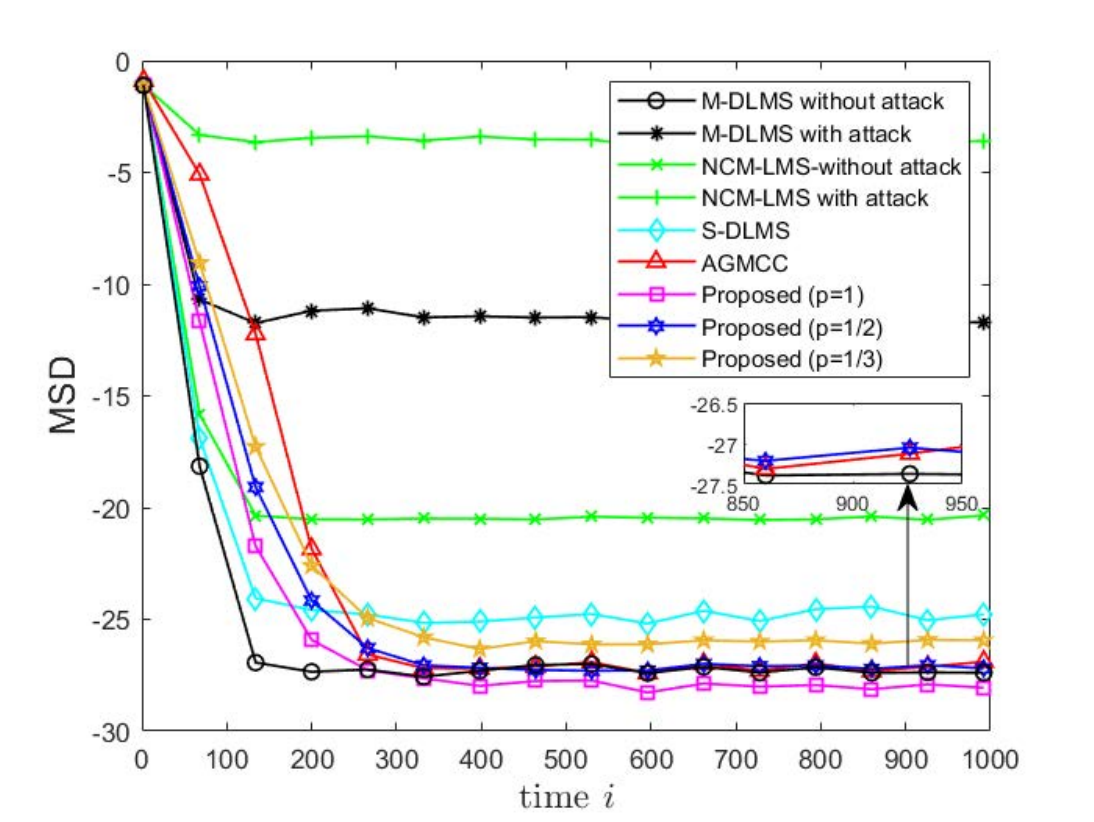}%
	}\hspace{0.1in}%
	\subfloat[Link attack]{%
		\includegraphics[width=0.34\textwidth]{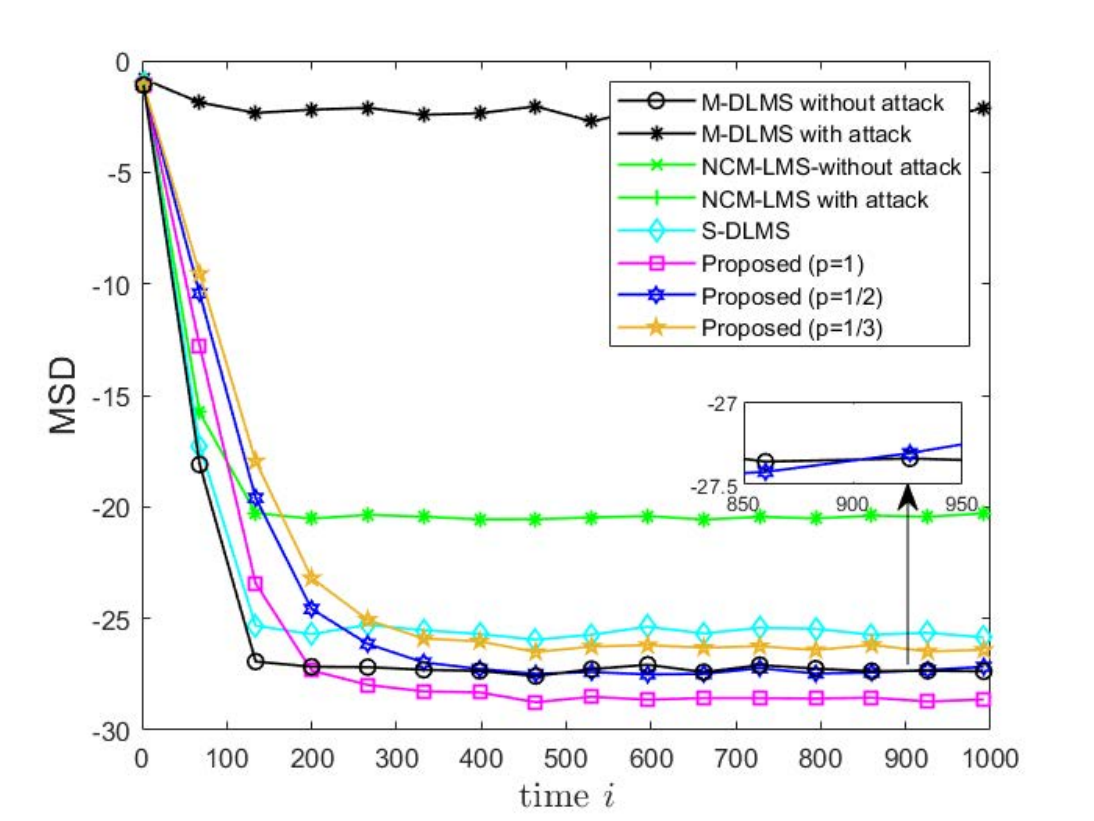}%
	}
	\caption{The MSD performance of different algorithms.
		(a) FDI attack. (b) Link attack.}
	\label{experiment}
\end{figure}

In the experiment, the attacker randomly select three nodes from the set ${\{2, 3, 6, 8, 13\}}$ for the attack. As shown in Fig. \ref{experiment}, the proposed  algorithm outperforms other algorithms including the M-DLMS algorithm when the communication load is the same. This is because, the proposed algorithm has more up-to-date data that can be used to train more refined models in the later stable stage. The proposed algorithm still exhibits comparable performance to M-DLMS when $p=\frac{1}{2}$. Finally, it is worth noting that when the communication nodes are less than half, the proposed algorithm continues to operate normally without being affected by attacks present in the network. However, nodes need to update the W-SVDD model frequently to counteract the impact of outdated and abnormal data, as shown in Table \ref{table1}.

\section{Conclusion}
  A low-communication resilient distributed estimation algorithm based on memory mechanism is proposed in this brief. The proposed algorithm reduces communication load through a node selection method, utilizes the W-SVDD method for detecting anomalous data, introduces an event-triggered mechanism to minimize ineffective model updates, and alleviates computational burden. Simulation results indicate that despite an increase in storage overhead, the proposed algorithm outperforms related work, either achieving better performance or matching the performance of other algorithms with half the communication load.

\bibliographystyle{IEEEtran}
\bibliography{refer}

\end{document}